\documentclass[12pt, prd, showpacs]{revtex4}
\usepackage{amssymb}
\usepackage{amsmath}

\input{tcilatex}

\begin{document}

\title{Rapidly rotating spacetimes and collisional super-Penrose process}
\author{O. B. Zaslavskii}
\affiliation{Department of Physics and Technology, Kharkov V.N. Karazin National
University, 4 Svoboda Square, Kharkov 61022, Ukraine}
\affiliation{Institute of Mathematics and Mechanics, Kazan Federal University, 18
Kremlyovskaya St., Kazan 420008, Russia}
\email{zaslav@ukr.net }

\begin{abstract}
We consider generic axially symmetric rotating spacetimes and examine
particle collisions in the ergoregion. The results are generic and agree
with those obtained in the particular case of the rotating Teo wormhole in
N. Tsukamoto and C. Bambi, Phys. Rev. D \textbf{91}, 104040 (2015). It is
shown that for sufficiently rapid rotation, the energy of \ a particle
escaping to infinity can become arbitrary large (so-called super-Penrose
process). Moreover, this energy is typically much larger than the center-of
mass energy of colliding particles. In this sense the situation differs
radically from that for collisions near black holes.
\end{abstract}

\maketitle

\section{Introduction}

Interest to high energy collisions of particles in the strong gravitational
field significantly increased after the observation made in \cite{ban} (it
also renewed interest to more early works on the closed subject \cite{pir1}
- \cite{pir3}).\ It was shown there that if two particles collide near the
extremal Kerr black hole, the energy in the centre of mass $E_{c.m.}$ can
grow unbound provided one of particles has fine-tuned parameters. This
provoked a large series of works which extended this result to many other
scenarios with unbounded $E_{c.m.}$. Not going in details, all these
scenarios can be classified in terms of three independent parameters. The
first one is the proximity to the horizon (plus necessity of fine-tuning of
one of particles).\ It is measured by the value of the lapse function $N\,\ $%
in\ the point of collision. This is just the scenario of collisions
suggested in \cite{ban} or its direct generalizations.\ Moreover, the actual
horizon can be absent (say, a system is only on the threshold of its
formation). The second relevant characteristics is the angular momentum $L$
of one of particles. If it is large and negative, $E_{c.m.}$ can be also
large. This is scenario of collisions inside the ergoregion, not necessarily
close to the horizon (such a scenario was suggested in \cite{ergo} and
generalized in \cite{mergo}).

Quite recently, a new type of scenario was considered in \cite{wh1}. The
authors considered particle collisions near the throat of the Teo wormhole 
\cite{teo}. It turned out that for sufficiently small throat areal radius, $%
E_{c.m.}$ can be large. One can ask, what is the underlying reason of high $%
E_{c.m.}$ for this problem and whether the results are due to specific
features of the wormhole model or have a general character. The answer is
found in \cite{wz} where it was shown that the effect of unbound $E_{c.m.}$
stems from high values of the parameter $\omega $ responsible for rotation
taken in the point of collision. In this sense, the effect is universal and
applies to rapidly rotating wormholes and, in principle, to relativistic
stars as well. (Although its astrophysical relevance remains unclear.) Thus
we obtain the third option of getting unbound $E_{c.m.}$.

Up to now, we discussed the sources of high $E_{c.m.}$ Meanwhile, there is
one more aspect - the properties of debris that reach flat infinity after
collision. Their energy~$E$ is related to $E_{c.m.}$ in a rather indirect
way. One can ask - whether or not the efficiency of the energy extraction to
infinity may also become unbound? (Following \cite{card}, we will use the
term "super-Penrose process".) We can exploit the same classification in
terms of three parameters $(N,L,\omega )$. (As is already known, there
exists a counterpart to the high energy collisions that arises due to an
electric charge \cite{jl} instead of rotation but we do not discuss this
issue here. We also do not consider the influence of the electromagnetic
field on the energy extraction that can, in principle, increase efficiency
of the Penrose process \cite{nar}). As far as small $N$ is concerned, it
turned out that for collisions near black holes, amplification of the energy
is restricted \cite{p} - \cite{z} if both initial paricles moved along
ingoing trajectories. If one of particles is on an outgoing trajectory, the
energy $E$ can be quite large \cite{shnit}, \cite{pir152} or even formally
unbound \cite{card}, \cite{mod} but in the latter case very special
conditions are required\ and it is hard to satisfy them \cite{pir15}, \cite%
{epl1}. Due to recent results \cite{pn}, the super-Penrose process is
possible for the case of naked singularity in collisions where also small $N$
in the point of collision is required. Another possibility of the
super-Penrose process, that arises due to large negative $L$ of one of
particles was found in \cite{sergo}.

In the present work, we examine the third remaining possibility and consider
collisions with large $\omega $. We will see that this can be considered as
a potential source of the super-Penrose process and has an universal
character. We do not pretend to any astrophysical applications (that is
another separate issue) and only demonstrate that this version of the
super-Penrose process is possible in principle. In the particular case of
the Teo wormhole \cite{teo}, the results for the collisional Penrose process 
\cite{wh2} are reproduced.

\section{Basic equations}

We consider the metric of the form%
\begin{equation}
ds^{2}=-N^{2}dt^{2}+g_{\phi }(d\phi -\omega dt)^{2}+\frac{dr^{2}}{A}%
+g_{\theta }d\theta ^{2}\text{.}  \label{met}
\end{equation}

We restrict ourselves to the motion in the equatorial plane $\theta =\frac{%
\pi }{2}$. Then, we can always redefine the radial coordinate in such a way
that $N^{2}=A$. It is implied that the metric coefficients do not depend on $%
t$ and $\phi $. Correspondingly, the Killing energy $E_{0}=-mu_{0}$ and the
angular momentum $L=mu_{\phi }$ are conserved. Here, $m$ is the particle's
mass, $\frac{dx^{\mu }}{d\tau }$ is the four-velocity, $\tau $ is the proper
time. We assume that there is no horizon, so $N$ vanishes nowhere.

The equations of motion along the geodesics read%
\begin{equation}
m\dot{t}=\frac{X}{N^{2}}\text{,}  \label{t}
\end{equation}%
\begin{equation}
m\dot{\phi}=\frac{L}{g_{\phi }}+\frac{\omega X}{N^{2}}\text{,}  \label{phi}
\end{equation}%
\begin{equation}
X=E-\omega L\text{,}  \label{xel}
\end{equation}%
where dot denotes the derivative with respect to $\tau $. From the
normalization condition 
\begin{equation}
u_{\mu }u^{\mu }=-1  \label{norm}
\end{equation}%
we obtain that the radial momentum%
\begin{equation}
m\dot{r}=\sigma Z\text{,}  \label{z}
\end{equation}%
where 
\begin{equation}
Z=\sqrt{X^{2}-N^{2}(m^{2}+\frac{L^{2}}{g_{\phi }})}\text{,}  \label{Z}
\end{equation}%
$\sigma =\pm 1$ depending on the direction of motion. The forward-in-time
condition $\dot{t}>0$ entails%
\begin{equation}
X>0\text{.}  \label{x}
\end{equation}

\section{Centre of mass energy for head-on collisions}

Let particles 1 and 2 collide. One can define the centre of mass energy $%
E_{c.m.}$ according to%
\begin{equation}
E_{c.m.}^{2}=-P_{\mu }P^{\mu }\text{,}
\end{equation}%
where $P^{\mu }=m_{1}u_{1}^{\mu }+m_{2}u_{2}^{\mu }$ is the total momentum
in the point of collision, subscripts label particles. Then,%
\begin{equation}
E_{c.m.}^{2}=m_{1}^{2}+m_{2}^{2}+2m_{1}m_{2}\gamma \text{,}  \label{CM}
\end{equation}%
where $\gamma =-u_{1\mu }u_{2}^{\mu }$ is the Lorentz factor of relative
motion.

For simplicity, we assume now that%
\begin{equation}
L_{1}=L_{2}<0\,,  \label{1=2}
\end{equation}

\begin{equation}
m_{1}=m_{2}=m\text{, }E_{1}=E_{2}\geq m\geq 0\text{,}  \label{cond}
\end{equation}%
\begin{equation}
m_{3}=m_{4}=\mu \text{.}  \label{mm}
\end{equation}

For the definiteness, let us consider head-on collision, so $\sigma
_{1}=-\sigma _{2}$. Then, it follows from (\ref{t}) - (\ref{z}) that

\begin{equation}
\gamma m^{2}=\frac{X_{1}^{2}+Z_{1}^{2}}{N^{2}}-\frac{L_{1}^{2}}{g_{\phi }}=%
\frac{2\beta }{N^{2}}-m^{2}\text{,}  \label{ga}
\end{equation}%
where%
\begin{equation}
\beta =X_{1}^{2}-\frac{N^{2}L_{1}^{2}}{g_{\phi }}\text{.}  \label{bex}
\end{equation}%
It is seen that%
\begin{equation}
\beta \geq \frac{m^{2}}{2}N^{2}\geq 0\text{,}  \label{be}
\end{equation}%
\begin{equation}
E_{c.m.}^{2}=\frac{4\beta }{N^{2}}.  \label{cm}
\end{equation}%
Obviously, 
\begin{equation}
E_{c.m.}\geq 2\mu \text{.}
\end{equation}%
Then, using (\ref{cm}) we obtain the bound%
\begin{equation}
\mu ^{2}\leq \frac{\beta }{N^{2}}\text{.}  \label{bound}
\end{equation}

\section{Conservation laws in the point of collision}

The conservation of the energy and angular momentum in the collision gives us%
\begin{equation}
E_{1}+E_{2}=E_{3}+E_{4}  \label{e}
\end{equation}%
\begin{equation}
L_{1}+L_{2}=L_{3}+L_{4}\,.  \label{l}
\end{equation}%
As a consequence, we have%
\begin{equation}
X_{1}+X_{2}=X_{3}+X_{4}\text{.}
\end{equation}%
The conservation of the radial momentum reads

\begin{equation}
Z\equiv \sigma _{1}Z_{1}+\sigma _{2}Z_{2}=\sigma _{3}Z_{3}+\sigma _{4}Z_{4}.
\label{rad}
\end{equation}%
Condition (\ref{x}) for particle 3 gives us $L_{3}\leq \frac{E_{3}}{\omega }$%
. For particle 4 with the conservation laws (\ref{e}), (\ref{l}) taken into
account, it follows from the same condition that%
\begin{equation}
L_{3}\geq \frac{E_{3}-X_{1}-X_{2}}{\omega }\text{.}  \label{l>}
\end{equation}

\section{Scenario 1: identical particles with nonzero angular momenta}

In what follows, we assume that $\sigma _{3}=+1$ (particle 3 moves to
infinity) and, similarily to \cite{wh2}, examine this process for two
identical particles assuming (\ref{1=2}) - (\ref{mm}) .

We obtain from (\ref{l>}) that%
\begin{equation}
\frac{E_{3}-2E_{1}}{\omega }+2L_{1}\leq L_{3}\leq \frac{E_{3}}{\omega }\text{%
.}
\end{equation}%
Let us consider the collision with $\sigma _{1}=-\sigma _{2}$. Then, as
particles are identical, the total radial momentum is equal to zero and $%
\sigma _{4}=-\sigma _{3}=-1.$ Thus we have%
\begin{equation}
X_{3}^{2}-N^{2}(\mu ^{2}+\frac{L_{3}^{2}}{g_{\phi }})=X_{4}^{2}-N^{2}(\mu
^{2}+\frac{L_{4}^{2}}{g_{\phi }})\text{.}  \label{34}
\end{equation}%
Equations (\ref{l}), (\ref{e}) now read%
\begin{equation}
2L_{1}=L_{3}+L_{4}\text{,}  \label{2L}
\end{equation}%
\begin{equation}
2E_{1}=E_{3}+E_{4}.  \label{2E}
\end{equation}%
We want to obtain large positive $E_{3}$, so $E_{4}$ should be large
negative. This is possible in the ergoregion only, so we take%
\begin{equation}
g_{00}=-N^{2}+\omega ^{2}g_{\phi }>0\text{.}  \label{erg}
\end{equation}%
After some algebra, one finds from (\ref{34}), (\ref{2L}), (\ref{2E}) that%
\begin{equation}
L_{3}=\frac{P}{Q}\text{, }  \label{L3}
\end{equation}%
\begin{equation}
X_{3}=\frac{K}{Q}\text{,}
\end{equation}%
where%
\begin{equation}
P=X_{1}E_{3}-\beta \text{, }Q=\omega E_{1}-\frac{L_{1}g_{00}}{g_{\phi }}%
\text{.}  \label{pq}
\end{equation}%
\begin{equation}
K=E_{3}L_{1}\frac{N^{2}}{g_{\phi }}+\omega \beta \text{.}  \label{K}
\end{equation}%
Here,%
\begin{equation}
Q>0\text{, }E_{1}>0\text{, }L_{1}<0\text{, }g_{00}>0\text{.}
\end{equation}%
One can show that 
\begin{equation}
K>0,  \label{k}
\end{equation}%
so indeed $X_{3}>0$ in agreement with the forward-in-time condition. (The
proof of this important property is done in Appendix.)

The condition $Z_{3}^{2}\geq 0$ gives us%
\begin{equation}
X_{3}^{2}\geq N^{2}(\mu ^{2}+\frac{L_{3}^{2}}{g_{\phi }}).  \label{pos}
\end{equation}%
By plugging (\ref{L3}), (\ref{pq}) into (\ref{pos}) we obtain the inequality
for $E_{3}$:%
\begin{equation}
E_{3}^{2}-2E_{3}E_{1}-\beta \frac{g_{\phi }(\omega ^{2}-\frac{N^{2}}{g_{\phi
}})}{N^{2}}+\frac{g_{\phi }Q^{2}}{\beta }\mu ^{2}\leq 0\text{.}
\end{equation}%
Its solution turns out to be rather simple:%
\begin{equation}
E_{\pm }=E_{1}\pm s\text{,}
\end{equation}%
\begin{equation}
s=\frac{Q\sqrt{g_{\phi }}}{N}\sqrt{B}\text{, }B=1-\frac{\mu ^{2}N^{2}}{\beta 
}\text{,}  \label{s}
\end{equation}%
\begin{equation}
E_{-}\leq E_{3}\leq E_{+}\text{.}
\end{equation}%
As the Lorentz factor $\gamma >0$, it is seen from (\ref{bound}) that $B\geq
0$.

The energy extraction is measured by the characteristics%
\begin{equation}
\eta =\frac{E_{3}}{E_{1}+E_{2}}\text{.}
\end{equation}%
For the maximum possible energy $E_{+}$,%
\begin{equation}
\eta _{+}=\frac{1}{2}+\frac{\sqrt{B}\sqrt{g_{\phi }}}{2N}\frac{Q}{E_{1}}=%
\frac{1}{2}+\frac{\sqrt{B}\sqrt{g_{\phi }}}{2N}(\omega +\frac{\left\vert
L_{1}\right\vert g_{00}}{E_{1}g_{\phi }})\text{.}  \label{extr}
\end{equation}%
In the above formulas, all quantities are taken in the point of collision $%
r=\rho $. In partiuclar, $\omega =\omega (\rho )$.

\subsection{Rapid rotation}

Let us change parameters of the metric in such a way that, formally, 
\begin{equation}
\omega \rightarrow \infty  \label{inf}
\end{equation}%
in the point of collision. This is not to be confused with the behavior at
inifnity since, in principle, this is consistent with the condition 
\begin{equation}
\lim_{r\rightarrow \infty }\omega =0  \label{zero}
\end{equation}%
for given values of parameters. A particular example from which this
consistency can be seen is the Teo wormhole (see below in Sec. IX).

To avoid subtleties connected with the asymptotically nonflat spacetimes, in
what follows we will assume that (\ref{zero}) is valid, so the energy at
infinity has a clear and unambiguous meaning. Meanwhile, generalization to
the metric that is not flat at infinity is also possible.

Then, in (\ref{erg}), it is the second term that dominates, provided that $%
g_{\phi }$ does not tend to zero too rapidly, so in the point of collision%
\begin{equation}
\omega ^{2}g_{\phi }\gg N^{2}\text{,}  \label{con}
\end{equation}

\begin{equation}
g_{00}\approx \omega ^{2}g_{\phi }\rightarrow \infty ,  \label{00}
\end{equation}%
\begin{equation}
\beta \approx \omega ^{2}L_{1}^{2}\text{,}  \label{blarge}
\end{equation}%
\begin{equation}
E_{+}\approx \frac{\omega ^{2}\left\vert L_{1}\right\vert \sqrt{g_{\phi }}}{N%
}\sqrt{B},
\end{equation}%
$X_{1}=E_{1}+\omega \left\vert L_{1}\right\vert \rightarrow \infty ,$ 
\begin{equation}
L_{3}\leq L_{3}(E_{+})\approx \omega \left\vert L_{1}\right\vert \frac{\sqrt{%
g_{\phi }}}{N}\sqrt{B}\text{.}  \label{Llarge}
\end{equation}

Thus the extraction can become unbounded (as long as backreaction of
particles on the metric is neglected), $\frac{E_{+}}{E_{1}}\gg 1$ when%
\begin{equation}
\omega ^{2}\gg \frac{E_{1}N}{\left\vert L_{1}\right\vert \sqrt{g_{\phi }}%
\sqrt{B}}\text{.}
\end{equation}

Then, it follows from (\ref{bex}), (\ref{cm}) that 
\begin{equation}
E_{c.m.}\approx \frac{2\omega \left\vert L_{1}\right\vert }{N}\text{,}
\end{equation}%
so 
\begin{equation}
\frac{E_{+}}{E_{c.m.}}\approx \frac{\sqrt{g_{\phi }}\sqrt{B}\omega }{2}\gg 1
\label{1cm}
\end{equation}%
due to (\ref{con}), (\ref{00}).

In contrast to the situation with collisions near black holes where, as a
rule, $E_{+}=O(1)$ and either $E_{c.m.}=O(1)$ or $E_{c.m.}$ is unbound, here
the situaiton is opposite: $E_{+}$ becomes much larger than $E_{c.m.}\,!$

\section{Scenario 2: identical particles with zero angular momenta}

It is seen from the previous formulas describing asymptotic behavior of the
energies for large $\omega $ that there is a special case that deserves
separate treatment. Let

\begin{equation}
L_{1}=0=L_{2}\text{.}
\end{equation}%
Then, the asymtotic evaluaitons from the previous section do not apply but
general formulas remain valid and are simplified. We have from (\ref{xel}), (%
\ref{Z}), (\ref{bex}), (\ref{bound}), (\ref{L3}), (\ref{pq}), (\ref{s}) that 
\begin{equation}
Z^{2}=E^{2}-N^{2}m^{2}\text{, }X=E\text{, }\beta =E_{1}^{2}\text{,}
\label{bee}
\end{equation}%
\begin{equation}
\mu ^{2}\leq \frac{E_{1}^{2}}{N^{2}}\text{,}  \label{me}
\end{equation}%
\begin{equation}
L_{3}=\frac{E_{3}-E_{1}}{\omega },
\end{equation}%
\begin{equation}
s=\frac{\omega E_{1}\sqrt{g_{\phi }}}{N}\sqrt{B}\text{, }B=1-\frac{\mu
^{2}N^{2}}{E_{1}^{2}}\text{,}  \label{sb}
\end{equation}%
\begin{equation}
E_{\pm }=E_{1}\pm \frac{\omega \sqrt{g_{\phi }}}{N}\sqrt{E_{1}^{2}-\mu
^{2}N^{2}},  \label{+}
\end{equation}%
\begin{equation}
\eta _{+}=\frac{1}{2}+\frac{\omega \sqrt{g_{\phi }}}{2N}\sqrt{1-\frac{\mu
^{2}N^{2}}{E_{1}^{2}}}.
\end{equation}%
In doing so, it follows from (\ref{cm}), (\ref{bee}) that 
\begin{equation}
E_{c.m.}^{2}=\frac{4E_{1}^{2}}{N^{2}}\text{.}  \label{cm2}
\end{equation}%
It is seen from (\ref{+}), (\ref{cm2}) that in the limit $\omega \rightarrow
\infty $ the energy $E_{+}$ grows but $E_{c.m.}$ remains finite.

Thus the ratio%
\begin{equation}
\frac{E_{+}}{E_{c.m.}}\rightarrow \infty 
\end{equation}%
for both types of scenarios considered above (with $L_{1}\neq 0$ and $L_{1}=0
$). If $g_{\phi }$ remains finite, $\frac{E_{+}}{E_{c.m.}}=O(\omega )$ for
both types of scenarios. However, if $g_{\phi }\rightarrow 0$ in this limit,
the growth of this ratio is more slow. Such a situation happens just in the
case of the Teo wormhole \cite{teo} where on the throat $\omega =O(b^{-3})$
and $g_{\phi }=O(b^{2})$ \cite{wh1}, \cite{wh2} (see also below).

For the scenario under discussion, the mass of the escaping particle is
restricted according to (\ref{me}). There is no similar restriction for the
previous type of scenario since the right hand side of (\ref{bound}) grows
as $\omega ^{2}$ for large $\omega $ according to (\ref{blarge}).

\section{Causality, rapid rotation and the role of ergoregion}

In our previous consideration the parameter $\omega $ is sent formally to
infinity or, at least, becomes very large. On the first glance, this causes
violation of causality since in the metric (\ref{met}) the term $g_{\phi
}(\omega dt)^{2}\,\ $seems to dominate. Then, the interval apparently
becomes spacelike that is impossible for any massive particle traveling with
a speed less than a speed of light. Had this "obvious" conclusion been
correct, this would have made all the above results as well as those of the
previous works \cite{wh1}, \cite{wh2} on this subject physically
meaningless. Fortunately, this is not so.

First of all, one of essential ingredients in derivation of formulas is the
normalization condition (\ref{norm}) that clearly tells us that a world line
of \ particle is timelike, as it should be. But the question remains because
of seeming contradiction described in the previous paragraph. To explain the
issue under discussion, one is led to take into account the nature of a
region in which we discuss our scenarios, i.e. the properties of the
ergoregion. Outside it, there exist trajectories on which a particle can
remain at rest. Then, one can neglect $d\phi $ in (\ref{met}), obtain the
spacelike interval and say that the limiting transition $\omega \rightarrow
\infty $ is not legitimate.

However, inside the ergosphere the situation is qualitatively different.
Meanwhile, it is the region (\ref{erg}) that we are interested in. It is the
requirement of the timelike nature of a world line that makes a particle to
rotate. It follows from (\ref{met}) and the condition $ds^{2}<0$ that (say,
for a circle orbit $r=const$)%
\begin{equation}
\omega -\frac{N}{\sqrt{g_{\phi }}}<\frac{d\phi }{dt}<\omega +\frac{N}{\sqrt{%
g_{\phi }}}\text{.}  \label{fw}
\end{equation}

On the border of the ergoregion (\ref{erg}), the left hand side vanishes, so
inside this region a particle must rotate. Eq. (\ref{fw}) is well known and
can be found in textbooks.\ We listed it here just because of our context in
which we discuss the limiting transition $\omega \rightarrow \infty $. If
eq. (\ref{fw}) is satisfied, one can take this limiting transition safely.

It is also instructive to look at the dynamic characteristics for the
situation in question. According to eq. (15) of ref. \cite{k}, there exist
he relation between the energy and angular momentum of a particle%
\begin{equation}
X=E-\omega L=\frac{mN}{\sqrt{1-V^{2}}}\text{.}  \label{el}
\end{equation}

Here, $V$ is the velocity measured locally by the ZAMO (zero angular
momentum observer \cite{72}). The left hand side of (\ref{el}) should be
positive in view of (\ref{x}). Then, if $L<0\,\ $and $\,N=O(1)$, we can take
the limit $\omega \rightarrow \infty $. In doing so, 
\begin{equation}
V\approx 1-\frac{m^{2}N^{2}}{2\omega ^{2}L^{2}}\text{,}  \label{v1}
\end{equation}%
so $V\rightarrow 1$ approaches the speed of light from below. Thus
simultaneously $\frac{d\phi }{dt}\approx \omega $ can be indefinitely large
but, at the same time, $V$ remains slightly less than 1, so causality is
respected.

\section{Rapid rotation: wormholes versus black holes}

It is instructive to compare the possibility of fast rotation for black
holes and wormholes. Then, the counterpart of $\omega _{thr}$ (subscript
"thr" denotes quantities calculated on the throat of a wormhole) is the
angular velocity of the horizon $\omega _{H}$. For the Kerr metric, 
\begin{equation}
\omega _{H}=\frac{a}{2Mr_{+}},  \label{kerr}
\end{equation}%
\begin{equation}
R=2M,
\end{equation}%
where $M$ is the black hole mass, $a=\frac{J}{M}\leq M$, $J\leq M^{2}$ is
the angular momentum, the horizon radius $r_{+}=M+\sqrt{M^{2}-a^{2}}$, $R$
is the quantity of $\sqrt{g_{\phi \phi }}$ on the horizon in the equatorial
plane $\theta =\frac{\pi }{2}$. One can construct the quantity%
\begin{equation}
v_{e}=R\omega _{H}  \label{bh}
\end{equation}%
that can be considered as a counterpart of the corresonding quantity for
wormholes introduced in eq. (7) of \cite{sc}.

Then, $v_{e}=\frac{a}{r_{+}}$, so the equation 
\begin{equation}
v_{e}\leq 1  \label{ve}
\end{equation}%
is satisfied.

In \cite{sc}, where rotating wormholes supported by a phantom scalar field
were considered, it was assumed that 
\begin{equation}
v_{e}\equiv \omega _{thr}R\leq 1  \label{7}
\end{equation}%
is satisfied, where now $R=(\sqrt{g_{\phi \phi }})_{thr}$.

Meanwhile, there is a big difference between both cases. For wormholes of
Ref. \cite{sc}, the mass-angular momentum relation reads%
\begin{equation}
M=2\omega _{thr}J\text{.}  \label{smarr}
\end{equation}

Formally, it looks like the Smarr mass formula for the case of extremal
black holes \cite{sm}, if $\omega _{thr}$ is replaced with $\omega _{H}$.
But for such black holes $J=M^{2}$, so we obtain that $\omega _{H}=\frac{1}{%
2M}$ in agreement with (\ref{kerr}). Then, $v_{e}=1$. Thus for a given mass $%
M$ both $J$ and $\omega _{H}$ are fixed. Meanwhile, for the wormhole \cite%
{sc} it follows from (\ref{smarr}) that $\omega _{thr}$ $=\frac{M}{2J}$,
where for a given $M$ we have freedom in changing $J$

If a black hole is nonextremal,%
\begin{equation}
J<M^{2}\text{,}
\end{equation}%
so in (\ref{bh})$\ v_{e}<1$.

The relevance of the inequality $J\leq M^{2}$ follows from the existence of
a black hole horizon. But for wormholes there is no horizon and there is no
such a restriction. Correspondingly, one should consider $M,J$ and $\omega
_{thr}$ as three parameters restricted by one relation (\ref{smarr}) only,
so we can take, say, $\omega _{thr}$ as a free parameter. And, we can send
it formally to infinity, without restriction (\ref{7}).

Then, the quantity $v_{e}\rightarrow \infty $. It is certain that it has
nothing to do with superluminal speed and causality violation. Outside the
ergoregion, the quantity $v_{e}$ could be interpreted as a linear velocity
of a particle with $r=const$ with respect to an observer at rest. But the
throat for a big $\omega $ is located inside the ergosphere. In that region,
such interpretation is impossible for the reasons explained above. Any
observer rotates there. The simplest choice of an observer is the ZAMO one.
Then, he is himself comoving with respect to this rapidly rotating throat.
Instead, one may ask about the velocity of any geodesic particle with
respect to the ZAMO frame. But, according to (\ref{v1}), it does not exceed
the speed of light.

Thus the quantity $v_{e}$ inside the ergoregion does not have direct
physical meaning. Moreover, this is so even in the black hole context. For
example, although for extremal black holes $v_{e}=1$, as explained above, in
the ZAMO frame the velocity of a particle on a marginally bound near-horizon
orbit around the near-extremal Kerr hole has $V=1/\sqrt{2}\neq v_{e}$ - see
page 356 of \cite{72} and generalization in \cite{excirc}.

In our context, fast rotation implies the validity of \ eq. (\ref{con}).
Taking $N=O(1)$, we obtain for wormholes of \cite{sc} the condition $\omega
_{thr}R\gg 1$ or $J\ll MR$. In this sense, for rapid rotation the angular
momentum should be small. This unusual behavior can be attributed to the
fact that the solutions considered in \cite{sc} are necessarily rotating.
For a finite $\omega _{thr}$ and $J\rightarrow 0$ the mass $M\rightarrow 0$
as well. If, instead, we want to keep $M$ nonzero even for small $J$, we are
led to compensate the lack of angular momentum by increasing $\omega _{thr}$%
. Further discussion of properties of the solutions \cite{sc} is beyond the
scope of the present paper.

\section{Comparison with the results for the Teo wormhole}

The scenarios with collisions of identical particles considered above are
quite general in that the results apply to any metrric of the form (\ref{met}%
). It is instructive to trace how the particular results for the Teo
wormhole \cite{wh2} can be reproduced. For this wormhole \cite{teo} the
metric coefficients in the equatorial plane read

\begin{equation}
N=1\text{, }\omega =\omega (r,a)=\frac{2a}{r^{3}}\text{, }g_{\phi }=r^{2}%
\text{.}  \label{teo}
\end{equation}

Then, our eqs. (\ref{L3}) for the angular momentum with (\ref{pq}) taken
into account, turns into eq. (4.22) of \cite{wh2} where one should put $r=b$
at the throat. Eq. (\ref{extr}) for the maximum extraction efficiency
corresponds to eq. (4.26) of \cite{wh2}. The bound (\ref{bound}) coincides
with (4.11) of \cite{wh2}. In a similar way, there is agreement in other
formulas.

It is seen from (\ref{teo}) that simultaneously 
\begin{equation}
\lim_{b\rightarrow 0}\omega (b,a)=\infty ,
\end{equation}
\begin{equation}
\lim_{r\rightarrow \infty }\omega (r,a)=0\text{,}
\end{equation}%
so rapid rotaiton in the region $r\sim b$ is compatible with the asymptotic
flatness!

\section{Scenario 3: collision of different particles with finite angular
momenta of debris}

In this Section, we relax restriction (\ref{1=2}). We also assume that 
\begin{equation}
L_{3}=O(1)\text{, }L_{4}=O(1)\text{ }
\end{equation}%
and do not fix the relation between $\sigma _{1}$ and $\sigma _{2}$
beforehand. The scenario under consideration does not coincide with previous
two cases where $L_{3}$ was big for big $\omega $ according to (\ref{Llarge}%
). To avoid cumbersome expressions we consider at once large values of $%
\omega $ which we are interested in. To ensure the conditions (\ref{x}) for
particles 1 and 2, the angular momenta of initial particles should satisfy%
\begin{equation}
L_{1}<0,L_{2}<0\text{.}  \label{12}
\end{equation}%
The case when one of angular momenta is equal to zero is also possible but
for simplicity we restrict ourselves by (\ref{12}) only.

Then, for $\omega \rightarrow \infty $, it follows from (\ref{Z}), that 
\begin{equation}
Z_{i}\approx X_{i}-\frac{N^{2}}{2X_{i}}(m_{i}^{2}+\frac{L_{i}^{2}}{g_{\phi }}%
)  \label{z12}
\end{equation}%
for each particle, $X$ is given by (\ref{xel}).

We expect%
\begin{equation}
E_{3}=\omega L_{0}+O(1),
\end{equation}%
where the unknown quantity $L_{0}=O(1)$. The conservation of the radial
momentum (\ref{rad}) gives us in the main approximation with respect to $%
\omega $, when only the leading terms are retained in (\ref{z12}) that%
\begin{equation}
-\sigma _{1}L_{1}-\sigma _{2}L_{2}=L_{0}-L_{3}+\sigma _{4}(-L_{0}-L_{4})%
\text{,}  \label{sl}
\end{equation}%
where we took into account that $\sigma _{3}=+1$.

The forward-in-time condition (\ref{x}) for particles 3 and 4 gives rise to%
\begin{equation}
L_{3}<L_{0}<-L_{4}\text{.}  \label{034}
\end{equation}

We must consider different variants separately, depending on signs of radial
momenta in the point of collision (further, the sign of $\sigma _{4}$ can
change if particle 4 bounces from the turning point). Let us denote
corresponding configurations as $(\sigma _{1},\sigma _{2},\sigma _{4})$.
Then direct inspection with (\ref{l}), (\ref{12}) and (\ref{034}) taken into
account shows that such configurations are possible: ($+,+;$ $+$), $(+,-,-)$
and $(-,+,-)$. In the first case eq. (\ref{sl}) turns into identity, so for
finding $L_{0}$ one should take into account the corrections in in the
expansion (\ref{z12}). In the second case, one finds that $L_{0}=L_{3}-L_{1}$
and in the third one $L_{0}=L_{3}-L_{2}$. By contrast with two previous
scenarios, in the second and third cases $E_{3}$ has the same order $\omega $
as $E_{c.m.}$.

\section{Discussion and conclusions}

We generalized observations concerning the properties of the collisional
Penrose process that were made in \cite{wh2} for the Teo wormhole. We
revealed the underlying reason of large unbound $E$ (the super-Penrose
process) and demonstrated \ that this happens due to large $\omega $ in the
point of collision and has a universal character. Thus we, in a sense,
filled a gap in logical classification of the cases of the super-Penrose
process and, in addition to the previously considered in literature types of
scenarios with small $N$ and large negative $L$, explored also possibilities
which are open due to large $\omega $. Thus in combination with previous
works \cite{epl1}, \cite{sergo} all potential factors of unbound $E_{c.m.}$
are examined in the "triad" ($N$,$L$,$\omega $). Although we implied that
the spacetime approaches the Minkowskian one at infinity, so that the notion
of the Killing energy is defined unambiguously, it is clear from derivation
that our approach admits, in principle, extension to more general asymptotic
behavior of the metric as well.

It is worth noting that the results apply to the region where $\omega $ is
big irrespective of whether or not the metric is asymptotically flat.
Meanwhile, both conditions ($\omega $ is big inside the ergosphere but $%
\omega \rightarrow 0$ at infinity) are compatible with each other as it was
demonstrated for the Teo wormhole.

It was noticed in \cite{wh2} that the center-of mass energy of colliding
particles is not related directly to the energy extraction. Now, we saw
that, moreover, an interesting feature consists in that there exist
scenarios in which $E_{3}\gg E_{c.m.}$ In particular, high $E$ are
consistent with modest $E_{c.m.}$ for $L_{1}=L_{2}=0$ that generalizes the
corresponding scenario for the Teo wormhole \cite{wh2}. Meanwhile, the
aforementioned inequality can be valid in a more general case of nonzero $%
L_{1,2}$. In another type of scenario, one of angular momenta of outgoing
particles is taken to be finite by assumption. Then, $E_{3}=O(E_{c.m.})$ is
large.

Thus it seems that the super-Penrose process is quite typical for rapidly
rotating spacetimes although its details can be different for different
scenarios.

As far as the set of objects to which our results can apply is concerned,
one can call, in the first place, rotating wormholes. In recent years,
interest to such spacetimes increased. Some general properties of slow
rotating wormholes were studied in \cite{sp1}, \cite{sp2}. Exact solutions
were obtained in \cite{sc} with the use of both analytical and numerical
methods. Rotating wormholes were also investigated analytically in \cite{rcw}
(although asymptotically their metric does not have the Minkowskian form).
Our results show that nontrivial effects \ due to particle collisions occur
not only for the Teo wormhole \cite{wh2} but for any kind of a wormhole
provided it rotates sufficiently rapidly. The question about potential
applications to relativistic stars is less clear. Formally, the results of
the present work are valid for them as well. The problem, however, is that
for known stars it is difficult to satisfy condition (\ref{con}).

Meanwhile, we would like to stress that our goal consisted not in the
investigation of some concrete properties of particular astrophysically
relevant models.\ Instead, we tried to reveal all potential possibilities of
the super-Penrose process as a general phenomenon in gravitation and fill
some gaps left after studies in physics of black holes \cite{pir152}, \cite%
{pir15}, \cite{epl1}, \cite{mas}, \cite{pos} and naked singularities \cite%
{pn}. Our results have a model-independent character and, in this sense,
show some interesting general properties of spacetime in the context related
to particle collisions. It is the generality of the results that is their
main advantage. What is especially interesting is that relation between the
energy in the centre of mass frame and that measured at infinity can be
quite different (that generalizes the corresponding observation made in \cite%
{wh2} about the Teo wormhole).

Nonetheless, careful examination of different star models is also of
interest that requires, however, a separate treatment.

\begin{acknowledgments}
This work was funded by the subsidy allocated to Kazan Federal University
for the state assignment in the sphere of scientific activities.
\end{acknowledgments}

\section{Appendix}

Here, we prove inequality (\ref{k}). As $L_{1}\leq 0$, $K\geq K_{+}$ where $%
K_{+}$ corresponds to the maximum possible value of $E_{3}=E_{+}+s$.
Therefore, it is sufficient to show that $K_{+}>0$. Using (\ref{K}), (\ref%
{bex}), (\ref{s}) one finds%
\begin{equation}
K_{+}=a_{0}+a_{1}L_{1}+a_{2}L_{2}^{2}\text{,}  \label{k+}
\end{equation}%
where%
\begin{equation}
a_{0}=\omega E_{1}^{2}>0\text{,}
\end{equation}%
\begin{equation}
a_{1}=\frac{N^{2}E_{1}}{g_{\phi }}(1+\frac{\omega \sqrt{g_{\phi }B}}{N}%
)-2E_{1}\omega ^{2}\text{.}
\end{equation}%
\begin{equation}
a_{2}=\frac{g_{00}}{g_{\phi }}(\omega -\frac{N\sqrt{B}}{\sqrt{g_{\phi }}}).
\end{equation}%
By assumpion, the point of collision is located inside the ergoregion, where
(\ref{erg}) is satisfied, hence%
\begin{equation}
\omega \sqrt{g_{\phi }}>N\text{.}  \label{omn}
\end{equation}

It is seen from (\ref{sb}) that $B<1$. Then,%
\begin{equation}
a_{2}>\frac{g_{00}}{g_{\phi }}(\omega -\frac{N}{\sqrt{g_{\phi }}})>0.
\end{equation}

The coefficent $a_{1}$ can be rewritten as 
\begin{equation}
a_{1}=E_{1}c\text{,}
\end{equation}%
where%
\begin{equation}
c=\frac{N^{2}}{g_{\phi }}-2\omega ^{2}+\frac{N}{g_{\phi }}\omega \sqrt{%
g_{\phi }B}.
\end{equation}%
Using the inequality $B<1$ again,%
\begin{equation}
c\leq \frac{N^{2}}{g_{\phi }}-2\omega ^{2}+\frac{N}{\sqrt{g_{\phi }}}\omega .
\end{equation}

It follows from (\ref{omn}) that%
\begin{equation}
c\leq \frac{N^{2}}{g_{\phi }}-\omega ^{2}=-\frac{g_{00}}{g_{\phi }}<0.
\end{equation}

Thus, since $L_{1}\leq 0$, all terms in (\ref{k+}) are positive or, at
least, non-negative. This completes the proof.

\end{document}